\documentclass{emulateapj}
\usepackage{graphicx}
\usepackage{natbib}
\usepackage{subfigure}

\shorttitle{Dynamics of $z>1$ spheroidals}
\shortauthors{Newman \etal}

\begin{document}

\newcommand{\etal}{et~al.}
\newcommand{\msol}{\,\textrm{M}_\sun}                
\newcommand{\lsol}{\,\textrm{L}_\sun}                
\newcommand{\hst}{\textit{HST}}
\newcommand{\mdyn}{M$_{\rm dyn}$}
\newcommand{\mstar}{M$_{*}$}

\title{Keck Spectroscopy of $z>1$  Field Spheroidals: Dynamical Constraints
on the Growth Rate of Red ``Nuggets"}
\author{Andrew B. Newman,\altaffilmark{1}
Richard S. Ellis,\altaffilmark{1}
Tommaso Treu,\altaffilmark{2}
and
Kevin Bundy\altaffilmark{3,4}}
\email{anewman@astro.caltech.edu}
\altaffiltext{1}{Department of Astronomy, California Institute of Technology,
Pasadena, CA 91125}
\altaffiltext{2}{Department of Physics, University of California, Santa Barbara, CA 93106-9530; Packard Fellow}
\altaffiltext{3}{Hubble Fellow}
\altaffiltext{4}{Astronomy Department, University of California, Berkeley, CA 94705}

\begin{abstract}
We present deep Keck spectroscopy for 17 morphologically-selected
field spheroidals in the redshift range $1.05<z<1.60$ in order to
investigate the continuity in physical properties between the claimed
massive compact red galaxies (``nuggets'') at $z\simeq2$  and
well-established data for massive spheroidal galaxies below
$z\simeq1$.  By combining Keck-based stellar velocity dispersions with
\hst-based sizes, we find that the most massive systems (M$_{\rm
dyn}>10^{11}$ M$_\odot$) grew in size over $0<z<1.6$ as
(1+z)$^{-0.75\pm0.10}$ (i.e., $\times 2$ since $z = 1.5$) whereas intermediate mass systems
($10^{11} {\rm M}_\odot > {\rm M}_{\rm dyn} > 10^{10} {\rm M}_\odot$) did not grow
significantly. These trends are consistent with a picture in which
more massive spheroidals formed at higher redshift via ``wetter''
mergers involving greater dissipation. To examine growth under the
favored ``dry'' merger hypothesis, we also examine size growth at a
fixed velocity dispersion. This test, uniquely possible with our
dynamical data, allows us to consider the effects of ``progenitor
bias.'' Above our completeness limit ($\sigma > 200$ km~s$^{-1}$), we find
size growth consistent with that inferred for the mass-selected
sample, thus ruling out strong progenitor bias. To maintain
continuity in the growth of massive galaxies over the past 10 Gyr, our
new results imply that size evolution over $1.3<z<2.3$, a period
of 1.9~Gyr, must have been
even more dramatic than hitherto claimed if the red sources at $z>2$
are truly massive and compact.
\end{abstract}
\section{Introduction}

The observation that many red galaxies with large stellar masses at
$z\simeq2$ are 3--5 times more compact than equivalent ellipticals in
the local Universe \citep[e.g.,][]{Daddi2005,Trujillo2007,vanDokkum2008,Buitrago2008,Damjanov2009}
has been a source of much puzzlement. How can an early
galaxy grow primarily in physical size without accreting significant
stellar mass as required if these objects are the precursors of the
most massive ellipticals observed today? Furthermore, studies of the
fundamental plane and other stellar population indicators do not
permit substantial recent star formation since $z\sim2$ in massive galaxies, thus
precluding growth by accretion of young stars or via gas-rich (``wet")
mergers \citep[e.g.,][hereafter T05]{Treu2005}. Some have questioned the reliability
of the observations, suggesting an underestimate of physical sizes or
an overestimate of stellar masses (\citealt{Hopkins2009}; however, see
\citealt{Cassata2009} for a contrasting view). Others have proposed size
expansion driven by self-similar dissipationless ``dry'' mergers, or
mass accretion from minor mergers \citep[][and references therein]{Khochfar2006,Naab2009,Hopkins2010}.

To verify the compact nature of distant sources and to track their
evolution in size and mass, it is preferable to use dynamical masses
\mdyn\, from absorption line spectra, which do not suffer from uncertainties
associated with the assumed initial mass function and stellar mass
estimates derived from broad-band photometry \citep[e.g.,][]{Muzzin2009}.
\mdyn\ measurements are available for relatively large
samples out to $z\sim1$ (T05; \citealt{vanderWel2008}, hereafter vdW08),
suggesting a small but detectable difference in average size at fixed
mass when compared to the local universe. But beyond $z\simeq1$, there
is little high-quality dynamical data for field spheroidals. \citet{vanDokkum2009}
undertook an heroic observation of a single $z>2$
source with a stellar mass $\simeq 2 \times 10^{11}M_\odot$ and an effective
radius $r_e=0.8$ kpc typical of compact galaxies at $z\simeq2.3$. The
spectrum has a claimed stellar velocity dispersion of
$\sigma=510^{+165}_{-95}$ km~s$^{-1}$, suggesting a remarkably dense
system. van Dokkum et al.~postulate the initial dissipative collapse at
$z\simeq3$ of a high mass ``core'' but are unable to account for its
subsequent evolution onto the $z\simeq1$ scaling relations. The
quantitative effect of minor mergers on the physical size of a galaxy
involves many variables, and it is unclear whether such dramatic size
evolution is possible while maintaining the tightness of the
fundamental plane and its projections \citep{Nipoti2009}.

Interpretation of the observed trends at fixed \mdyn\, is further
complicated by the so-called ``progenitor bias'' \citep{vanderWel2009}:
if galaxies grow by dry mergers, the main progenitor of a
present-day massive galaxy did not have the same mass at
$z\sim2$. Similarly, if galaxies become recognizable as spheroidals
only above a certain threshold in stellar velocity dispersion
$\sigma_{\rm ET}$ that depends on redshift, it is clear that the
addition of a new -- and less dense -- population could mimic a false
evolutionary trend. This bias can be reduced by considering galaxy
sizes at fixed $\sigma$.  Foremost, $\sigma$ changes very little under
a variety of growth mechanisms \citep[e.g.,][]{Hopkins2010} and it is
therefore a better ``label'' than \mdyn~to track the assembly
history. Secondly, $\sigma$ is closely correlated with stellar age
\citep{vanderWel2009} and
therefore offers the most direct way to track the evolving population.

Given there is no clear consensus in understanding the continuity
between the galaxy population at $z<1$ and that at $z>2$, we have
embarked on a campaign to measure $\sigma$ and \mdyn for a large
sample of field spheroidals at $1<z<1.7$. This has recently become
practical using multi-object optical spectrographs equipped with deep
depletion red-sensitive CCDs. Our goal is to extend the earlier work
at $z<1$ (T05, vdW08) to within $\simeq$1 Gyr of the
sample of ultracompact galaxies at $z\simeq2.3$. In this first
analysis, we present new results spanning the redshift range
$1.05<z<1.60$.

We adopt a $\Lambda$CDM cosmology with $(\Omega_m, \Omega_v, h) =
(0.3, 0.7, 0.7)$; all magnitudes are in the AB system. A Chabrier IMF
is assumed where necessary.

\section{Sample and Observations}

Our targets were selected from archival \emph{HST}/ACS data in the EGS (GO 10134,
PI: Davis), SSA22 (GO 9760, PI: Abraham \& GO 10403, PI: Chapman), and GOODS-N
(PI: Giavalisco) fields. For the EGS, we used the \citet{Bundy2006} catalog which
matches CFHT ($BRI$, \citealt{Coil2004}; $ugriz$, CFHTLS) and Palomar ($JK_s$)
photometry. Photometric redshifts are supplemented by spectroscopic
redshifts from the DEEP2 survey. For SSA22, we used
a photometric redshift catalog based on Subaru ($BVRIz$) and UH~2.2m
($JHK_s$) imaging kindly provided by P. \citet{Capak2004}. In GOODS-N, we used
the \citet{Bundy2009} catalog which matches ACS and Subaru $K_s$ photometry. Galactic
extinction corrections were based on the dust maps of
\citet{Schlegel1998}. The parent sample for spectroscopic study in EGS and SSA22 was
defined by $I-K_s > 2$, $I < 23.5$, and $z > 1$; in GOODS-N, the photometric criteria
were ${\rm F850LP} - K_s > 1.5$ and ${\rm F850LP} < 23.5$.
All galaxies satisfying these criteria were visually inspected in the ACS images by
one of us (RSE) and those with E/S0 or early-disk morphology retained.

Keck I LRIS observations were made for 14 EGS and SSA22 targets on
2009 June 26--28 in median seeing of $0\farcs9$. The
600~mm${}^{-1}$ grating blazed at 1~$\mu$m was used, providing a
velocity resolution of $\sigma_{\rm inst} = 58$~km~s${}^{-1}$ at
9000~\AA. The total integration times were 40.8~ks and 32.4~ks in the
EGS and SSA22 fields, respectively. On 2010 April 5--6 LRIS observations were made
of 7 GOODS-N targets with 34.8~ks of integration in $0\farcs8$ seeing. One additional
GOODS-N spectrum was secured with Keck II DEIMOS observations on 2010 April 11--12
using the 831~mm${}^{-1}$ grating. The LRIS data were reduced using the
code developed by \citet{Kelson2003}. Spectra were extracted using
optimal weighting based on Gaussian fits to the spatial
profile. Telluric absorption correction and relative flux calibration
were provided by a DA star observed at matching airmass at the end of
each night.

\begin{figure}[t!]
\centering
\includegraphics[width=0.8\linewidth]{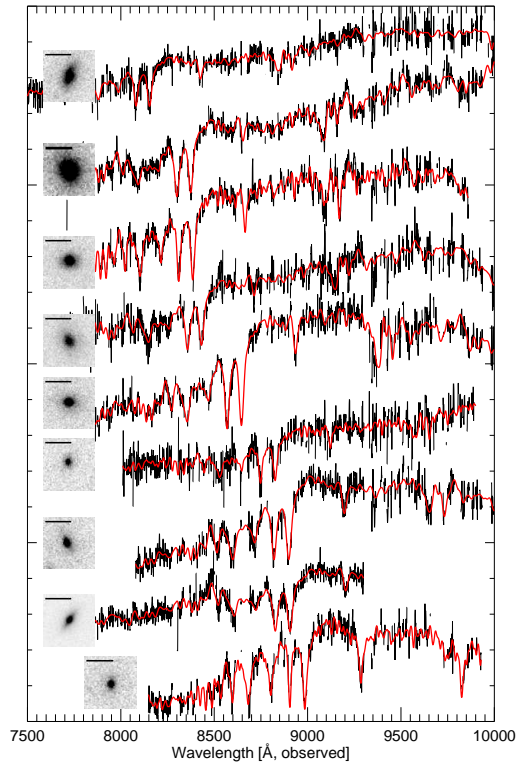}
\caption{Keck spectra of our 17 targets ordered by redshift (continued in Fig.~\ref{fig:spectra2}).
Each is smoothed with a 3 pixel (2.4~\AA) running median with sky lines
omitted (black) and compared to fits to broadened stellar templates
(red).  \hst~images are inset with a 1'' ruler. The order of objects matches
that in Table~1.\label{fig:spectra}}
\end{figure}
\begin{figure}[t!]
\centering
\includegraphics[width=0.8\linewidth]{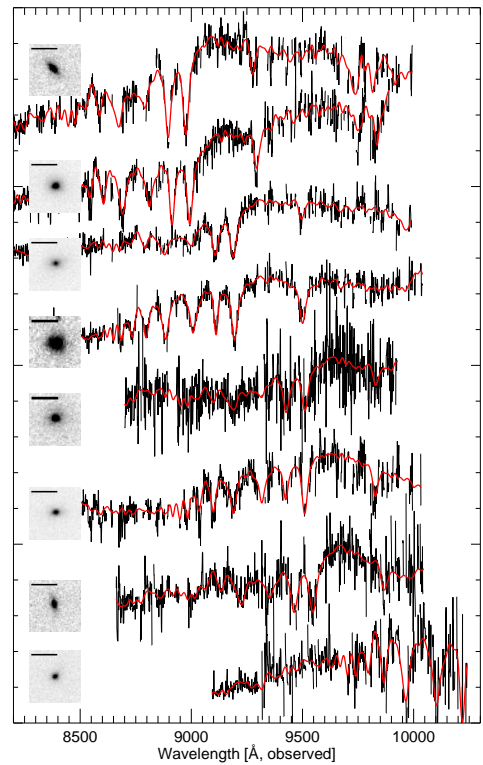}
\caption{Continuation of Figure~\ref{fig:spectra}.\label{fig:spectra2}}
\end{figure}


\section{Velocity Dispersions and Photometry}

We measured stellar velocity dispersions, $\sigma$, by fitting
broadened stellar templates using the PPXF code of
\citet{Cappellari2004}. The instrumental resolution was measured using
unblended sky lines; their variation with wavelength was well fit by a
low-order polynomial. The template collection comprised 348 stars of
type F0--G9 from the Indo-US coud\'{e} library \citep{Valdes2004} with
a range of metallicities and luminosities (classes III--V).
We verified that including A star templates does not affect our measurements.
For each galaxy, PPXF constructed an optimal template as a linear
combination of these stellar spectra, although our results do not
significantly differ if the best-fitting single template is used. To
avoid systematic errors, we masked pixels contaminated by OH emission.
Based on tests with the continuum filtering, sky masking threshold,
and stellar template choices, we assigned a systematic uncertainty to
each velocity dispersion, typically $5 - 10\%$. We were able to secure
a reliable dispersion for 17/22 galaxies (see Figs.~\ref{fig:spectra} and \ref{fig:spectra2}). Velocity
dispersions were corrected to an effective circular aperture of radius
$R_e/8$ as described in \citet{Treu1999}; the mean correction factor
is 1.13. 

Surface photometry was measured in the \hst~images using GALFIT
\citep{Peng2002} with a PSF determined from a nearby
isolated star. F814W imaging was used in EGS and SSA22, while F850LP data
were adopted in GOODS-N.
For consistency with the local SDSS sample, we fit de
Vaucouleurs profiles and determine circularized radii.  We also fit
S\'{e}rsic profiles but found that the mean S\'{e}ersic index $n$ is consistent with 4
(i.e., de Vaucouleurs).
We estimate uncertainties of $\sim10\%$ in $R_e$ based on testing the background
level, simulating the recovery of synthetic de Vaucouleur profiles
placed in blank sky patches, and comparing with the independent
measurements of vdW08 for the T05
subsample.

We convert the observed ACS magnitude to the rest $B$ magnitude by
matching the observed $I-K_s$ color to a grid of
\citet{BC03} single-burst models of varying age and
metallicity. The uncertainty in this $k$-correction is $\sim
4\%$. Based on the optical and NIR photometry discussed in \S2,
stellar masses were estimated using the Bayesian stellar
population analysis code developed by \citet{Auger2009}.
An exponentially-decaying star formation history
was assumed. Table~1 summarizes the dynamical and photometric properties for
our sample of 17 $z>$1 galaxies. 

Our data allow us to compare photometrically-derived stellar masses to
dynamical masses, which we define as $M_{\rm dyn} = 5 \sigma^2 R_e / G$.
Overall these are in good agreement:
$\langle \log M_{\rm dyn}/M_* \rangle = 0.17 \pm 0.07$, consistent with the difference
between local dynamical and stellar masses inferred by \citet{Cappellari2006}
and with the dark matter fractions (or heavier IMFs) found by independent methods
\citep[e.g.,][]{Graves2010, Treu2010}.

\begin{deluxetable*}{ccccccccccccc}
\tablecolumns{13}
\tablecaption{Photometric and Spectroscopic Data}
\tablehead{\colhead{ } & \colhead{R.A.} & \colhead{Dec.} & \colhead{Morph.} & \colhead{$z$} & \colhead{$R_e$} & \colhead{$\sigma$} & \colhead{M${}_B$} & \colhead{$I - K_s$} & \colhead{$z - K_s$} & \colhead{$K_s$} & \colhead{$\log M_{\rm dyn}/\msol$} & \colhead{$\log M_*/\msol$}}
\startdata
E1 & 214.9853 & 52.9513 & S0? & 1.054 & 6.44 & $228 \pm 32$ & -22.32 & 2.14 & 1.70 & 19.79 & $11.59 \pm 0.13$ & 11.17 \\
S1 & 334.3529 & 0.2734 & E & 1.110 & 4.74 & $242 \pm 18$ & -22.72 & 2.63 & 1.81 & 19.58 & $11.51 \pm 0.07$ & 11.23 \\
E2 & 214.9702 & 52.9911 & E/S0 & 1.113 & 4.02 & $151 \pm 15$ & -22.18 & 2.15 & 1.69 & 20.16 & $11.03 \pm 0.09$ & 10.91 \\
E3 & 215.0061 & 52.9755 & Sab & 1.124 & 6.11 & $266 \pm 28$ & -22.20 & 2.40 & 2.01 & 20.04 & $11.70 \pm 0.09$ & 11.01 \\
E4 & 214.9847 & 52.9614 & E/S0 & 1.179 & 2.65 & $258 \pm 19$ & -22.21 & 2.23 & 1.76 & 20.02 & $11.31 \pm 0.07$ & 11.00 \\
E5 & 214.9815 & 52.9501 & E & 1.225 & 1.43 & $139 \pm 25$ & -21.20 & 2.45 & 1.93 & 21.16 & $10.51 \pm 0.16$ & 10.57 \\
E6 & 215.0351 & 52.9830 & S0 & 1.243 & 2.38 & $302 \pm 29$ & -22.20 & 2.57 & 2.06 & 19.79 & $11.40 \pm 0.08$ & 11.19 \\
GN1* & 189.2681 & 62.2264 & Sab & 1.253 & 1.29 & $290 \pm 26$ & -22.19 & 2.79 & 2.07 & 20.26 & $11.10 \pm 0.08$ & 11.07 \\
E7$\dagger$ & 215.1319 & 53.0163 & E/S0 & 1.262 & 1.57 & $103 \pm 21$ & -21.83 & 2.58 & 2.02 & 20.43 & $10.29 \pm 0.18$ & 10.90 \\
E8 & 215.1371 & 53.0173 & Sab & 1.262 & 1.38 & $280 \pm 34$ & -21.90 & 2.38 & 1.90 & 20.76 & $11.10 \pm 0.11$ & 10.91 \\
GN2 & 189.0634 & 62.1623 & E/S0 & 1.266 & 1.58 & $239 \pm 36$ & -22.24 & 2.54 & 1.74 & 20.55 & $11.02 \pm 0.13$ & 10.88 \\
GN3 & 188.9345 & 62.2068 & E/S0 & 1.315 & 3.44 & $288 \pm 28$ & -22.32 & 2.77 & 2.07 & 20.45 & $11.52 \pm 0.09$ & 11.01 \\
S2$\dagger$ & 334.3502 & 0.3032 & Sab & 1.315 & 2.47 & $171 \pm 29$ & -22.37 & 2.29 & 1.51 & 20.80 & $10.93 \pm 0.15$ & 10.69 \\
S3 & 334.4233 & 0.2256 & E/S0 & 1.394 & 2.50 & $271 \pm 71$ & -22.01 & 3.20 & 2.44 & 20.74 & $11.33 \pm 0.23$ & 11.09 \\
GN4$\dagger$ & 189.1132 & 62.1325 & E & 1.395 & 0.77 & $206 \pm 49$ & -21.93 & 2.24 & 1.71 & 21.24 & $10.58 \pm 0.21$ & 10.55 \\
E9 & 215.1219 & 52.9576 & S0/a & 1.406 & 1.19 & $337 \pm 79$ & -22.02 & 2.39 & 2.19 & 20.82 & $11.20 \pm 0.21$ & 10.83 \\
GN5$\dagger$ & 188.9625 & 62.2286 & E & 1.598 & 0.68 & $274 \pm 41$ & -22.51 & 2.71 & 2.25 & 20.69 & $10.77 \pm 0.13$ & 11.02
\enddata
\tablecomments{$\dagger$ denotes systems with strong Balmer absorption. * denotes the DEIMOS observation. R.A.~and Dec.~are in the J2000 equinox. $R_e$ is in kpc. Velocity dispersions (in km~s$^{-1}$) are standardized to a circular aperture of radius $R_e/8$, and uncertainties include the systematic component added in quadrature. M$_B$ is the rest-frame $B$ band absolute magnitude. $I-K_s$ and $z-K_s$ photometry has been interpolated to a uniform system of Bessell $I$ and SDSS $z$. Dynamical mass uncertainties reflect the total uncertainty (statistical and systematic) in $\sigma$. Stellar masses assume a Chabrier IMF. The order of galaxies (by redshift) matches Figs.~\ref{fig:spectra} and \ref{fig:spectra2}.}
\end{deluxetable*}

\section{Size Evolution}

Size evolution has commonly been studied by comparing spheroids of the
same mass at different epochs. Although an unlikely evolutionary path
for individual galaxies, it is the most observationally convenient
approach, particularly at $z > 1.5$ where dynamical measures are
scarce. We first conduct this comparison in~\S~\ref{ssec:sizemass}.
Following our discussion in \S1, we then examine size evolution at
fixed velocity dispersion in \S~\ref{ssec:sizesigma}.

\subsection{Size evolution at fixed mass}
\label{ssec:sizemass}

We constructed a local reference sample for comparison purposes using 
SDSS DR7 spectroscopic data \citep{SDSSEDR}. Red sequence galaxies with $0.05 < z <
0.15$ were selected using a color cut \citep[Eq.~1 of][]{Yan2006}.
Velocity dispersions were corrected to a standard aperture of $R_e/8$
as in \citet{Jorgensen1995}.
The resulting mean dynamical mass--radius relation is
$R_e = 2.88~(M_{\rm dyn} / 10^{11} \msol)^{0.55}$~kpc, based on de
Vaucouleurs radii measured in SDSS and interpolated to rest $B$
band.\footnote{The relation is consistent with that of vdW08, taking
into account the different apertures to which we normalize velocity
dispersions.}

\begin{figure}
\plotone{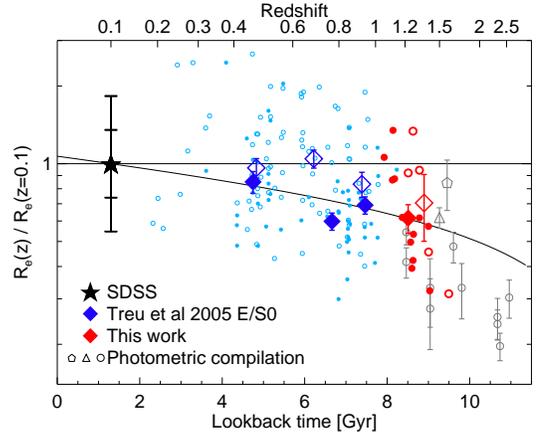}
\caption{Size evolution of spheroids at fixed mass.
Solid diamonds show the mean size and its uncertainty for
massive ($\log M_{\rm dyn}/\msol > 11$) spheroids relative to red sequence
galaxies in SDSS of the same dynamical mass (computed in log
space). Open diamonds refer to intermediate-mass ($10 < \log M_{\rm dyn}/\msol <
11$) systems. The solid line shows a $(1+z)^{-0.75\pm0.10}$
fit to the more massive sample. Individual galaxies in the spectroscopic samples
are shown by colored circles.
Light gray points with error bars are based on photometric stellar masses and show mean sizes relative to SDSS galaxies of the same stellar mass, based on the \citet{Shen2003} relation, for several samples: \citet[][triangle]{Saracco2010}, \citet[][pentagon]{Mancini2010}, and the vdW08 (circles) compilation.
\label{fig:size}}
\end{figure}

Figure~\ref{fig:size} shows the size evolution inferred by comparing the
present sample and the T05 E/S0 galaxies to this SDSS relation. 
The T05 and present samples are well-matched in morphology and rest optical colors 
and so comprise an excellent dataset for studying evolution over a wide
redshift interval.
A simple power law fit $(1+z)^{-0.75\pm0.10}$ to the $M_{\rm dyn} > 10^{11} \msol$
sample is indicated by the solid line. This corresponds to a 40\% decrease in
size by $z=1$, and marginally slower evolution than that inferred by
vdW08 for $M_{\rm dyn} > 3\times10^{10} \msol$ spheroidals
($(1+z)^{-0.98\pm0.11}$; corresponding to a 50\% decrease at
$z=1$). However, the difference is partly explained by the correction
applied by vdW08 to their measured sizes based on simulations. In
contrast, we do not apply any corrections to our measured sizes given
the lack of consensus on this matter in the literature \citep[see
also][]{Hopkins2009,Mancini2010} and the results of our own
simulations.

Our $\langle z \rangle = 1.3$ sample probes an epoch within 2 Gyr of
the $z\sim2.3$ samples whose compact sizes have motivated the present work.
Figure~3 illustrates that, given the size dispersion in the dynamical sample
at a given redshift and the difficulties of comparing our dynamical sample
with one whose masses are likely less precise, the sharp drop in size
seen over this short time interval may not be that significant.  
If confirmed, however, the implied size evolution is quite large compared to the
expected accretion from mergers over the same period, which we estimate to be 40\% of
a typical $10^{11} \msol$ galaxy at $z = 1.3$.\footnote{Estimated using
the merger rate calculator presented in \citet{HopkinsCalculator}.}
Thus, if minor mergers are responsible, these results imply a very high efficiency of
$d\log R_e / d \log M \simeq 2.6$ for growing spheroids, just consistent with the upper
end of estimates determined from merger simulations \citep[e.g.,][]{Hopkins2010}.

Also striking is the different
trend seen in less massive $10 < \log M_{\rm dyn}/\msol < 11$ galaxies (open
diamonds in Figure~\ref{fig:size}). Although the high-$z$
samples are small and include some compact examples, we find
no evidence for mean size evolution over $0 < z < 1.6$,
i.e.~$\propto(1+z)^{0.02\pm0.15}$. This is consistent
with a picture in which more massive galaxies formed earlier and from
wetter mergers with more dissipation, creating more compact remnant
spheroids \citep[e.g.,][]{Khochfar2006,Trujillo2006}, and at variance with the
model proposed by \citet{vanderWel2009} in which lower-mass galaxies are the most
strongly affected by progenitor bias and display the strongest evolution.
However, we caution that the lower mass samples may be affected by
selection effects, since the brighter -- and therefore possibly larger --
objects may be preferentially selected given our flux
limits. This is not a concern for the $>$10$^{11}$
M$_{\odot}$ sample, where we are complete for any reasonable
mass-to-light ratio.
A characterization of the bias requires a 
self-consistent model with Monte Carlo simulations, which is beyond the
scope of this Letter and is left for future work when larger samples will
be available.

\subsection{Size evolution at fixed velocity dispersion}
\label{ssec:sizesigma}

\begin{figure}
\plotone{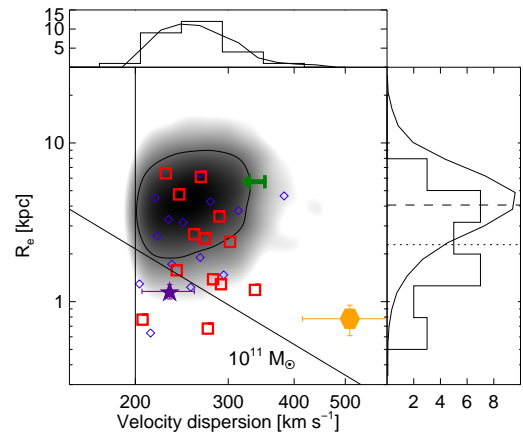}
\caption{Distribution of effective radii at matched velocity dispersion for 
spheroids with $\sigma > 200$~km~s${}^{-1}$ in the present sample
(red), the T05 $z>0.9$ sample (blue), and SDSS (smoothed grayscale
with $1\sigma$ contour).  The histogram (this data) and solid line
(SDSS) in the top panel agree by construction. The right panel
compares the distribution of $R_e$ for the matched samples. Dashed and
dotted lines indicate the means in log space. For comparison, the
$z\sim1.7$ stacked sample of \citet{Cappellari2009} is shown (star),
along with the single galaxies of
\citet[][hexagon, $z=2.186$]{vanDokkum2009} and \citet[][green upper limit, $z=1.823$]{Onodera2010}.
The diagonal line represents \mdyn=10$^{11}$ M$_\odot$.\label{fig:sigmaRe}}
\end{figure}

Comparisons at fixed mass may be affected by ``progenitor bias.'' A
preferred approach, when dynamical data are available, is to examine
galaxies of the same \emph{velocity dispersion}. This offers the two
advantages discussed in \S~1 and is illustrated in
Figure~\ref{fig:sigmaRe}. A cut in \mdyn\ includes only the largest
galaxies at a fixed $\sigma$. Therefore, if galaxies below some
threshold $\sigma_{\rm ET}$ are missing from the high-redshift
samples, this could mimic an evolutionary trend in mass-selected
samples. According to the preferred prescription of
\citet{vanderWel2009}, $\sigma_{\rm ET}=233$ km~s$^{-1}$ at
$z=1.3$. Therefore we should expect to see some effect for our sample,
even though the presence of lower $\sigma$ objects in our sample
already suggests that the progenitor bias is not as strong.

Figure~\ref{fig:sigmaRe} shows this comparison in terms of the
$\sigma-R_e$ plane. To avoid luminosity selection biases, we consider
only galaxies with $\sigma > 200$~km~s${}^{-1}$, where we are 90\%
complete, based on the SDSS distribution of luminosity at fixed
$\sigma$ and a conservative estimate of luminosity evolution,
consistent with passive evolution of an old stellar population formed
at $z_f=3$.  The SDSS galaxies (grayscale) are weighted so as to match
the $\sigma$ distribution of the $0.9 < z < 1.6$ sample (top panel).
The right panel of Figure~\ref{fig:sigmaRe} then compares the size
distributions of the local and $0.9 < z < 1.6$ samples at matching
$\sigma$. By fitting for size evolution at fixed $\sigma$ we find that
sizes evolve as $(1+z)^{-0.88\pm0.19}$. The good agreement with the
size evolution inferred at fixed mass rules out strong progenitor
bias.  A full evolutionary model with selection effects is needed to
include objects with lower $\sigma$ and quantify progenitor bias and
size evolution more accurately. This is left for future work with
larger samples.

\section{Conclusions}

Our Keck spectra have shown the utility of securing individual
spectroscopic and photometric measures for a representative sample of
$z>1$ massive spheroidals. By probing to $z\simeq1.6$, we are sampling
velocity dispersions, sizes and dynamical masses within 1.2 Gyr of the
puzzling population of compact red galaxies at $z \simeq 2.3$.

Importantly, the size evolution we infer over $0<z<1.6$ at fixed
dynamical mass is modest: $\simeq\times2$ for the most massive ($\log
M_{\rm dyn}/M_\odot>11$) examples but much smaller for lower mass systems. If
the compact red galaxies at $z\simeq2-2.3$ are their precursors, they
must have grown dramatically in size over a very short time interval.

\acknowledgments
We thank Peter Capak for kindly providing photometric catalogs for the
SSA22 field and Arjen van der Wel for helpful comments regarding size
measurements. It is a pleasure to acknowledge Matt Auger for computing the
stellar masses and Dan Stark for the DEIMOS observation. We thank
the anonymous referee for comments that improved the quality of this paper.
We are grateful to the staff of the Keck Observatory and Connie
Rockosi, in particular, for ensuring our early access to the upgraded
LRIS-R was successful. Research support by the Packard Foundation is
gratefully acknowledged by TT. 
The authors wish to recognize and acknowledge the cultural role and reverence
that the summit of Mauna Kea has always had within the indigenous Hawaiian community.
We are most fortunate to have the opportunity to conduct observations from
this mountain.
\vspace{1mm}

\bibliographystyle{apj}
\bibliography{ms}

\end{document}